\documentclass[12pt]{article}
\usepackage{epsf}
\usepackage{amsmath}
\usepackage{amsfonts}
\usepackage{amssymb}
\usepackage{fullpage}
\usepackage{graphicx}
\usepackage{color}
\usepackage{psfrag}
\begin{document}
%
%
\pagestyle{empty}
\rightline{IFT-UAM/CSIC-12-57}
\vspace{1.5cm}
\begin{center}
\LARGE{\bf Holographic Flow of \\ Anomalous Transport Coefficients
\\[12mm]}
\large{Karl Landsteiner\footnote{karl.landsteiner@uam.es} and Luis
Melgar\footnote{luis.melgar@csic.es}\\[3mm]}
\normalsize{ Instituto de F\'{\i}sica Te\'orica IFT-UAM/CSIC,\\[-0.3em] 
C/ Nicol\'as Cabrera 13-15 \\ 
Universidad Aut\'onoma de Madrid, 28049 Madrid, Spain} \\ [2mm]
\vspace*{3cm}
\small{\bf Abstract} \\[5mm]
\end{center}
We study the holographic flow of anomalous conductivities induced by gauge and
gravitational Chern-Simons terms.
We find that the contribution from the gauge Chern-Simons term gives rise to a
flow that can be interpreted
in terms of an effective, cutoff dependent chemical potential. In contrast the
contribution of the gauge-gravitational
Chern-Simons term is just the temperature squared and does not flow. 
%
%
\newpage
\pagestyle{plain}
\setcounter{page}{1}
\setcounter{footnote}{0}
\section{Introduction}
Anomalies appear in the context of relativistic quantum field theories. In four
dimensions chiral anomalies \cite{[a]} 
involve triangle diagrams with either only vector currents or vector currents
and the energy momentum tensor, in which case
one speaks of a (mixed gauge-) gravitational anomaly \cite{[b]}. They are
responsible for the breakdown of a
classical symmetry due to quantum effects.
If the symmetry is local anomalies impose severe restrictions on the
structure and definition of gauge theories (for comprehensive reviews on
anomalies see  \cite{[c]}).
In the case of a symmetry generated by $T_A$, and considering only right-handed
fermions, the presence of a chiral anomaly in vacuum is encoded in a
non-vanishing $d_{ABC} = \frac{1}{2} \mathrm{tr} \left(T_A\{T_B,T_C\}\right)$.
The corresponding parameter in the case of the gravitational anomaly is $b_A =
\mathrm{tr}\left(T_A \right)$.

Recently, it has been pointed out that at finite temperature and density,
anomalies are responsible for the appearance of new non-dissipative transport
phenomena \cite{[x],[y],[d]}. In the chiral magnetic effect an external magnetic
field induces a current parallel to
it

\begin{eqnarray}J^{\mu} = \sigma_B B^{\mu} \end{eqnarray}

where $\sigma_B$ is the chiral magnetic conductivity and $B^\mu = \frac 1 2 \epsilon^{\mu\nu\rho\lambda}u_\nu F_{\rho\lambda}$. 
A second effect is the chiral vortical effect.
It refers to the creation of a current parallel to vortices in the fluid

\begin{eqnarray}J^{\mu} = \sigma_V \omega^{\mu} \end{eqnarray}

with $\omega^{\mu}= \epsilon^{\mu \nu \rho \lambda}u_{\nu}
\partial_{\rho}u_{\lambda} $ being the vorticity vector and $u_{\mu}$ the
fluid four-velocity.

The contribution of the gravitational anomaly to these transport
coefficients was first obtained in a weakly coupled gas of chiral fermions
in \cite{[7b]}. A holographic model that confirmed these findings at strongly
coupling was developed and studied in \cite{[7]}.

The chiral magnetic and the chiral vortical conductivities can be calculated
from first principles via the Kubo formulae \cite{[5]} (Latin letters denote
purely spatial indices)
 
\begin{eqnarray} \label{sigma_B} \sigma_B = \lim_{p_n\rightarrow 0 }
\frac{i}{2p_c} \sum_{a,b} \epsilon_{abc} \left<J^aJ^b \right>(\omega=0, \vec p)
\\
\label{sigma_V} \sigma_V =\lim_{p_n\rightarrow 0 } \frac{i}{2p_c} \sum_{a,b}
\epsilon_{abc} \left<J^aT^b_0 \right> (\omega=0, \vec p) \end{eqnarray}

There are related transport coefficients for the energy  current $T^{0\mu}$ 
\begin{eqnarray}
{T}^{0\mu} &=& \sigma^\epsilon_B B^{\mu}\,, \\
{T}^{0\mu} &=& \sigma^\epsilon_V \omega^{\mu} \,.
\end{eqnarray}
They are calculated via the Kubo formulae 
\begin{eqnarray} \label{sigmaepsilon_B} \sigma^\epsilon_B = \lim_{p_n\rightarrow
0 }
\frac{i}{2p_c} \sum_{a,b} \epsilon_{abc} \left<T_0^aJ^b \right>(\omega=0, \vec
p)
\\
\label{sigmaepsilon_V} \sigma^\epsilon_V =\lim_{p_n\rightarrow 0 }
\frac{i}{2p_c} \sum_{a,b}
\epsilon_{abc} \left<T_0^aT^b_0 \right> (\omega=0, \vec p) \end{eqnarray}

The role played by the gravitational anomaly was further studied in \cite{[12]}, 
where an ideal Weyl gas in arbitrary even dimensions was
considered. This leads to a generalization of  the anomalous
conductivities valid for any (even) dimension and an expression that relates the
anomaly induced transport coefficients to the anomaly polynomial of the Ideal
Weyl gas. Furthermore, in \cite{[13]}, a definition for the local entropy
current for higher-curvature gravitational theories was proposed and the
Fluid/Gravity correspondence was applied to compute the first order
conductivities in the presence of the gravitational anomaly. \\

Within the gauge-gravity duality the running
with the holographic coordinate can be interpreted as a type of renormalization
group (RG) flow 
in the dual field theory \cite{[e]}. The first application of this holographic
flow to transport coefficients is \cite{[4]} where it was shown that the
electric conductivity
and the shear viscosity have a trivial flow.
It is known now that
some of the transport coefficients indeed present a non-trivial flow (see, for
instance \cite{[4b]}). The extension to finite chemical potential has been
studied in \cite{[4c]}.  \\
Recently, there is a renewed interest in this subject due to the explicit
holographic construction of the Wilsonian Renormalization Group \cite{[8]}, 
which has made possible to show that multi-trace
deformations in the effective action are induced after integrating out high
energy modes.  Finally, in \cite{[16]} it was
pointed out that all the apparently different frameworks used over time to study
the holographic flow are actually equivalent.\\
 
It is natural to analyze the holographic flow of the anomalous conductivities as
well. 
In this paper we present several
approaches to compute the
different flows and show that all the methods lead to the same
results as expected.  In Section 2 the setup is presented.
We interpret the holographic flow, defined as in \cite{[4]}, as a
cutoff flow that
arises by varying the holographic cutoff at finite holographic coordinate value
$r=\Lambda$. Section 3 studies the case of gauge fields without taking into
account the backreaction onto
metric perturbations.
Then we generalize the previous approach to include also the
metric perturbations and present a non-covariant method to calculate the
flow equations for the retarded Green's functions. Somewhat surprisingly we do
find
a non-trivial flow but give it a natural interpretation as a cutoff flow of an
effective chemical potential.
The flow of the
correlators is computed by explicitly solving the equations of motion for a
system restricted to live between the black brane horizon and a hyper surface
placed at finite $r=\Lambda$, which acts as a boundary.  The section
ends with a discussion regarding the compatibility of the results so obtained
with the flow equations. It is shown that both approaches
are in agreement. In Section 4 the attention is focused on the gravitational
anomaly. We discuss subtleties concerning the definition of a Dirichlet problem
and the necessity for the inclusion of a boundary term to ensure that the
correct form for the divergence of the current is found. Contrary to
what happens for the chemical potential, we find that the temperature term
stemming
from the gravitational anomaly does not flow.

We conclude in Section 5 with a discussion.

\section{Setup}

Lets show how transport coefficients flow with a variation of
the holographic cutoff scale. We define the theory with a cutoff as:

\begin{eqnarray}  
S = \frac{1}{e^2}\int_{r<\Lambda} \sqrt{-g}\left(-\frac{1}{4} F_{MN}F^{MN} \,.
\right) 
\end{eqnarray} 

We consider this theory in a general black brane background of the form

\begin{eqnarray} ds^2 = -g_{tt} dt^2 + g_{rr} dr^2 + g_{ii} d (x^i)^2\,.
\end{eqnarray}

We assume that the above metric has an event horizon at $r=r_H$ and that every
component depends only on $r$. The boundary is placed at $r =\Lambda$. The
metric
is also assumed to be regular except at the horizon and possibly in the limit
$\Lambda\rightarrow \infty$.
The current of the holographic dual field theory is

\begin{eqnarray} J^{\mu} = \left.\frac{1}{e^2}
\sqrt{-g} F^{\mu r}  \right|_{r=\Lambda} \,. \end{eqnarray}

In the gauge $A_r=0$ the $x$-component of its variation due to a small
perturbation of 
the gauge field reads

\begin{eqnarray} \label{eq:jx}
J^{x} = \left.\frac{-1}{e^2}
\sqrt{-g}g^{xx}g^{rr} \dot a(x,r) \right|_{r=\Lambda}\end{eqnarray}

where $\dot a  =da/dr$ is the r-derivative of the aforementioned perturbation.

We define $a(x,r)$ as $a(x,r) = \frac{a(r)}{a(\Lambda)}a^{(0)}(x)$, so that it
is normalized at the boundary to $a(x,\Lambda) = a^{(0)}(x)$ and $a(r)$ solves
the radial wave equation 

\begin{eqnarray}\ddot a(r) +\frac{1}{2}\dot a(r)\left(g^{tt}\dot g_{tt} +
g^{xx}\dot g_{xx} -g^{rr}\dot g_{rr}\right) + g_{rr}\left(\omega^2g^{tt}- 
k^2g^{xx}\right)a(r)=0\end{eqnarray} 

On the other hand, we \emph{define} the electric conductivity at the boundary as
$J^x = \sigma_E(\Lambda)E(\Lambda)$, where $E(\Lambda) =  -i\omega a^{(0)}$
is the external applied electric field. Comparing
this to equation (\ref{eq:jx}) we conclude 

\begin{eqnarray} \label{eq:sigmaE}
\sigma_E(\Lambda)= \left.\frac{-i}{e^2 \omega}
\sqrt{-g}g^{xx}g^{rr} \frac{\dot
a(r)}{a(\Lambda)}\right|_{r=\Lambda}\end{eqnarray}

Varying the cutoff $\Lambda \rightarrow \Lambda + d\Lambda$ we find the for the 
differential of the electric conductivity

\begin{eqnarray} \frac{d\sigma_E(\Lambda)}{d\Lambda} = \frac{-i}{e^2  \omega}
\left[   \frac{d}{dr}  \left( \sqrt{-g}g^{xx}g^{rr} 
\frac{ da(r)/dr}{a(r)}\right)\right]_{r=\Lambda}\end{eqnarray}

This equation shows that we can study the flow of the transport coefficients
with the cutoff reformulating it as the evolution with respect to the coordinate
$r$, by formally identifying $r$ with $\Lambda$.

We can use now the equation of motion for the perturbation $a(r)$ and the
definition of 
the conductivity (\ref{eq:sigmaE}) to derive the flow equation

\begin{equation}
 \frac{d\sigma_E(\omega,k)}{d\Lambda} = -i \omega\left[ \frac{e^2}{\sqrt{-g}}
g_{rr}g_{xx} 
\sigma_E^2 + \frac{\sqrt{-g}}{e^2}
g^{xx}\left(g^{tt} + \frac{k^2}{\omega^2} g^{xx}\right) \right] 
\end{equation}

This the flow equation first derived in \cite{[4]}. It can be solved by
demanding infalling boundary conditions on the horizon. In
particular the flow for the DC conductivity turns out to be trivial $\dot
\sigma_E =0$. In this case the electric conductivity 
is completely determined by its value on the horizon via the membrane paradigm
\begin{equation}
 \sigma_E(\Lambda) = \sigma_E(r_H) = \frac{1}{e^2}\,.
\end{equation}

\section{Flow of anomalous conductivities}

We will apply now the strategy outlined before to the anomalous transport
coefficients.
Two models will be considered. First we discuss a model in which we neglect the
backreaction of the gauge field fluctuations on the metric. We will study the
interplay
between two $U(1)$ symmetries which we call vector and axial ones. This allows
to model
the chiral magnetic and the chiral separation effect.
A second model will use only one anomalous $U(1)$ symmetry but we will also
include the
backreaction onto the metric. This allows to study also the flow of the chiral
vortical
conductivity and the flow of the anomalous transport coefficients related to the
energy current.

\subsection{Vector and Axial symmetries}
We will apply the aforementioned strategy to the chiral magnetic conductivity
\cite{[2]} .
Its proper definition requires the interplay between a vector like $U(1)$
symmetry
and an axial $U(1)$ symmetry. Holographic models have been investigated in
\cite{Yee:2009vw,Rebhan:2009vc,[1]}. The model allows for the definition of
the chiral magnetic conductivity and axial conductivities involving external
axial magnetic fields. 
Its action is given by \cite{[1]}

\begin{eqnarray} \nonumber  S=  \int \sqrt{-g}(-\frac{1}{4g_V^2}
F^V_{MN}F^{MN}_V - \frac{1}{4g_A^2} F^A_{MN}F^{MN}_A + \\
+ \frac{\kappa}{2} \epsilon^{MNPQR}A_M \left(F^A_{NP}F^A_{QR} + 3
F^V_{NP}F^V_{QR}\right)) \end{eqnarray}

where $V$ stands for 'vector' and $A$ for 'axial'. The Lagrangian
contains two Maxwell actions for vector and axial gauge fields and
a particular choice of Chern-Simons term. In what follows,
we will stick to the notation of \cite{[1]}; concretely, we define 
the epsilon symbol as $\epsilon(ABCDE) =  - \sqrt{-g} \epsilon^{ABCDE}$,
with $\epsilon(rtxyz)=1$ ($r$ corresponds to the fifth coordinate).

From the boundary term of this action, after perturbing both the axial and the
vector gauge fields, we obtain an 
expression for the boundary theory currents

\begin{align} \label{eq:vectorcurrent} J^{\mu} &= \left. \left(
\frac{1}{g^2_V} \sqrt{-g} F_V^{\mu r} + 6 \kappa \epsilon^{\mu \nu \rho \lambda}
A_{\nu} F^V_{\rho \lambda}\right) \right|_{r=\Lambda}\,, \\
\label{eq:axialcurrent} J_5^{\mu} &= \left. \left(
\frac{1}{g^2_A}
\sqrt{-g} F_A^{\mu r} + 2 \kappa \epsilon^{\mu \nu \rho \lambda} A_{\nu}
F^A_{\rho \lambda}  \right)\right|_{r=\Lambda} \,, \end{align}

where $\epsilon^{\mu \nu \rho \lambda} \equiv \epsilon^{r \mu \nu \rho
\lambda}$. The coefficients in front of the Chern-Simons terms are crucial to
ensure that the vector current is non-anomalous $D_\mu J_V^\mu=0$. 
The axial current is anomalous
$D_{\mu} J_5^{\mu} = - \frac{\kappa}{2} \epsilon^{\mu \nu \rho \lambda} \left(3
F^V_{\mu \nu} F^V_{\rho \lambda} + F^A_{\mu \nu} F^A_{\rho \lambda} \right)$ 
\cite{[1]}. Comparing with the standard result from the one loop triangle
calculation, we find $\kappa = -\frac{N_c}{24 \pi^2}$ for a dual strongly
coupled $SU(N_c)$ gauge theory for a mass less Dirac fermion in the fundamental
representation. Note also that both currents are invariant under vector gauge
transformations
but not under axial gauge transformations.\\

The equations of motion for the gauge fields are

\begin{eqnarray} \frac{1}{g_A^2}\nabla_N F^{NM}_A + \frac{3 \kappa}{2}
\epsilon^{MNPQR} (F^A_{NP}F^A_{QR} + F^V_{NP}F^V_{QR}) = 0 \\
 \frac{1}{g_V^2}\nabla_N F^{NM}_V + 3 \kappa \epsilon^{MNPQR} (F^A_{NP}F^V_{QR})
= 0 \end{eqnarray}

In order to study the flow of the conductivities with the fifth coordinate, we
will proceed as follows:

\begin{itemize}
 \item We introduce an axial and vector perturbation of the gauge fields
\begin{eqnarray} A_{M} = A^{(0)}_{M} +  a_M (y,t,r)\\
V_M = V^{(0)}_M +  v_M(y,t,t) \end{eqnarray}

We switch on perturbations only in the
$z$ and $x$--directions (transverse directions): $a_z(y,t,r)$, $v_z(y,t,r),a_x(y,t,r)$, $v_x(y,t,r)$

\item Since the Chern-Simons contribution to the current depends only
on the intrinsic gauge fields on the cutoff surface, its flow
is trivial. The non-trivial part of a possible flow is 
completely contained in the covariant currents
\begin{eqnarray} \label{curr12} J^{(1) x} = \left.\left(
\frac{1}{g^2_V} \sqrt{-g} F_V^{(1)x r} \right) \right|_{r=\Lambda}\\
\label{curr22} J_5^{(1)x} =  \left.\left(\frac{1}{g^2_A}
\sqrt{-g} F_A^{(1) x r} \right) \right|_{r=\Lambda} \end{eqnarray} 

\item We define our transport coefficients as the response to the perturbations
and in terms of the previously defined covariant currents as
\begin{align}
\label{eq:constjx}
 J^{(1)x} &= \sigma_{CME} \epsilon(rtxyz) F_{yz}^{(1)V} +
\sigma_{axial} \epsilon(rtxyz) F_{yz}^{(1)A} \,,\\
\label{eq:constjx5}
 J_5^{(1)x} &= \sigma_{axial} \epsilon(rtxyz) F_{yz}^{(1)V} +
\sigma_{55} \epsilon(rtxyz) F_{yz}^{(1)A} \,,
\end{align}

$\sigma_{axial}$ defines the vector current generated by an external axial
magnetic field. 
Observe that, in order not to have $F^{(1) xr}_{\{A,V\}}=0$
identically, one has to turn on the perturbations $a_x(y,t,r), v_x
(y,t,r)$. However, these do not play a role when studying the flow of the anomalous conductivities for they induce contributions that tend to zero in the low $\omega$,
low $k$ limit, very much as occurs in \cite{[4]}.  

\end{itemize}

The value of the background gauge fields is \cite{[1]}

\begin{eqnarray} A^{(0)}_0 = \alpha - \frac{\mu_5 r_H^2}{r^2} \\
V^{(0)}_0 = \gamma - \frac{\mu r_H^2}{r^2} \end{eqnarray}
The integration constants $\alpha$ and $\gamma$ can be fixed by
e.g. demanding that the gauge fields vanish on the horizon. In any case
the covariant currents do not depend on these integration constants. The
consistent
currents (\ref{eq:vectorcurrent}), (\ref{eq:axialcurrent}) do however depend on
them through the Chern-Simons currents.
For a discussion of this dependence see \cite{Rebhan:2009vc,[1]}.

The procedure consists of using the equations of motion to find the value of
$\partial_r \sigma$, where $\sigma$ a generic conductivity defined at some
hyper surface $r=\Lambda$.
In fact, we only need the equations of motion projected onto $x$ and the Bianchi
identity associated with the indices $(r,y,z)$ to obtain an expression of the
derivative with respect to $r$ of
the different transport coefficients. From the simple form of our metric it can
be seen that the vector normal 
to a hyper surface of $x$=constant reads $n_{\mu}^x= \sqrt{g_{xx}} (0,0,1,0,0)$. Hence, for
the vector gauge field we have

\begin{eqnarray} n^x_{M} \left[\frac{1}{g_V^2}\nabla_N F^{NM}_V + 3 \kappa
\epsilon^{MNPQR} (F^A_{NP}F^V_{QR})\right] = 0  \end{eqnarray}

Taking advantage of the relation $\nabla_N F^{NM} =
\frac{1}{\sqrt{-g}}\partial_N \left( \sqrt{-g}F^{NM} \right)$ and the definition
of the currents (\ref{curr12}) and (\ref{curr22}), we arrive at

\begin{eqnarray} \label{difcurr} \partial_r J^{(1)x} = -12 \kappa \sqrt{-g}
\epsilon^{rtxyz} \left(F^{A(0)}_{tr}F^{V(1)}_{yz} + F^{A(1)}_{yz}F^{V(0)}_{tr}
\right)  \end{eqnarray}

where we have neglected $F_V^{(1)tx}; F_V^{(1)yx}$ for these modes lead to
vanishing contributions in the low momentum and low frequency limit, as
mentioned before. Besides, we have carried out the contraction $
\epsilon^{xNPQR}F^A_{NP}F^V_{QR} =-4 \epsilon^{rtxyz} (F^A_{tr}F^V_{yz}+
F^A_{yt}F^V_{rz}+ F^A_{tz}F^V_{ry} + (A \leftrightarrow V))$.
The Bianchi identity to first order associated with indices $(r,y,z)$ reads 

\begin{eqnarray} \partial_r F^{\{V,A\}(1)}_{yz}  + \partial_y
F^{\{V,A\}(1)}_{zr} + \partial_z F^{\{V,A\}(1)}_{ry} = 0 \end{eqnarray}

Assuming $\partial_z F^{\{V,A\}(1)}_{ry} \sim g_{rr}g_{yy} \partial_z J^{y(1)} =
0$ we obtain 
\begin{eqnarray} \label{bianch} \partial_r F^{\{V,A\}(1)}_{yz} =
-\frac{g_{zz}g_{rr}}{\sqrt{-g}} g^2_{\{V,A\}} \partial_y J_{\{V,A\}}^{(1)z}
\end{eqnarray} 
Now, making use of these ingredients, the computation of $\partial_r \sigma$ is
immediate:

\begin{eqnarray} \label{defCME} \partial_r \sigma_{CME} = \lim_{\omega, k
\rightarrow 0} \left[\frac{\partial_r J^{x(1)}
}{\epsilon(rtxyz)F^{V(1)}_{yz}}-\frac{J^{x(1)}}{(\epsilon(rtxyz)F^{V(1)}_{yz})^2
}
\partial_r F^{V(1)}_{yz}\right]_{a_M=0} \end{eqnarray}

Plugging (\ref{difcurr}) and (\ref{bianch}) into (\ref{defCME}) we find, in
momentum space

\begin{eqnarray} \partial_r \sigma_{CME} = \lim_{\omega, k \rightarrow 0}
\left[12 \kappa F^{A(0)}_{tr} +i k \sigma_{CME} \frac{g_{zz}g_{rr}}{\sqrt{-g}}
g^2_{V} \frac{J^{(1)z}}{\epsilon(rtxyz)F^{V(1)}_{yz}}\right] \end{eqnarray}

Taking the limit $\omega, k \rightarrow 0$ and substituting $F^{V(0)}_{tr} = -
\partial_r A^{(0)}_0 = -2\frac{\mu_5 r_H^2}{r^3}$, we get the following flow equation
for the chiral magnetic conductivity

\begin{eqnarray} \partial_r \sigma_{CME} = -24 \kappa \frac{\mu_5 r_H}{r^3}
\end{eqnarray}

whose solution is  

\begin{eqnarray} \sigma_{CME} =C + 12 \kappa \frac{\mu_5 r_H^2}{r^2} \,.\end{eqnarray}

$C$ is an integration constant that we must fix. In order to do that, we
impose in-falling boundary conditions for the perturbations (or,
equivalently, regularity at the horizon $r=r_H$ \cite{[1]}). This in turn
implies that the fields must depend only on the combination $dv = dt +
\sqrt{\frac{g_{rr}}{g_{tt}}}dr$ \cite{[4]}. Therefore, in the $A_r =0 $ gauge,
we have

\begin{eqnarray} \partial_rA_x = \sqrt{\frac{g_{rr}}{g_{tt}}} \partial_t A_x
\,\,\, at\,\,r = r_H \end{eqnarray} 

This condition forces directly $J^{(1)x} (r=r_H)$ to be 

\begin{eqnarray} j^{(1)x} (r=r_H) \sim F^{(1)xr} (r=r_H) \sim E^i \end{eqnarray}

Imposing these infalling boundary conditions results therefore in a vanishing
chiral magnetic conductivity at the horizon for the covariant current
\footnote{Note that
the consistent currents might have non-vanishing chiral magnetic conductivity on
the horizon due
to the Chern-Simons contribution and depending on the value of the integration
constant $\alpha$.}
Thus the integration constant $C$ can be fixed simply by the condition

\begin{eqnarray} \sigma_{CME} (r=r_H) = 0 \rightarrow C = -12 \kappa \mu_5 =
\frac{N_c \mu_5}{2 \pi^2} \end{eqnarray}  

(Recall that $\kappa = -\frac{N_c}{24 \pi^2}$). \\

The
transport coefficients (\ref{eq:constjx}) and (\ref{eq:constjx5}) can be
calculated in an analogous way.

For the axial current the projected equation is

\begin{eqnarray} n_{xM} \left[\frac{1}{g_A^2}\nabla_N F^{NM}_A + \frac{3
\kappa}{2} \epsilon^{MNPQR} (F^A_{NP}F^A_{QR} + F^V_{NP}F^V_{QR})\right] = 0
\end{eqnarray}

which implies

\begin{eqnarray} \partial_r j^{(1)x}_5 = \sqrt{-g}12 \kappa \epsilon^{rtxyz}
\left(\frac{2 \mu_5}{r^3} F^{(1)A}_{yz} + \frac{2 \mu}{r^3} F^{(1)V}_{yz}
\right) \end{eqnarray}

The values of the conductivities at $r=\Lambda$ then read

\begin{eqnarray} \label{CME} \sigma_{CME} (\Lambda) =  \frac{N_c \mu_5}{2 \pi^2}
\left(1-\frac{r_H^2}{\Lambda^2} \right) \\
\sigma_{axial} (\Lambda) =\frac{N_c \mu}{2 \pi^2} \left(1-\frac{r_H^2}{\Lambda^2}
\right)\\
\sigma_{55} (\Lambda) = \frac{N_c \mu_5}{2 \pi^2} \left(1-\frac{r_H^2}{\Lambda^2}
\right)\end{eqnarray} 

As expected, the result in the limit $r \rightarrow \infty$ is precisely the one
obtained in \cite{[1]} using AdS/CFT techniques.

In view of the topological nature and the non-renormalization theorem for the
chiral magnetic
conductivity it is at first sight somewhat surprising to find a non-trivial
flow. 
This flow becomes however natural if we define the chemical potential in its
elementary way
as the energy needed to introduce one unit of charge into the ensemble. In the
holographic dual
this corresponds to bring a unit of charge from the boundary, now situated at
$r=\Lambda$ behind
the horizon. The energy difference between a unit of charge at the boundary and
a unit of charge at
the horizon is just given by $A_0(\Lambda) - A_0(r_H) = \mu(\Lambda)$. This
defines an effective
chemical potential in the theory equipped with the cutoff $\Lambda$. In fact the
definition of such
an effective chemical potential is natural even in field theory. If we have a
momentum cutoff of 
order $\Lambda$ we can localize a unit charge only inside a volume within a
radius of order $1/\Lambda$.
Thermalizing this unit of charge means spreading it out over the entire
ensemble. The
difference in energy between the
two configurations, the unit of charge localized within $1/\Lambda$  and spread
out over the ensemble 
again is the effective chemical potential.  

All the anomalous conductivities can therefore be expressed in the form
\begin{equation}
 \sigma(\Lambda) = \frac{N_c \mu(\Lambda)}{2\pi^2}\,.
\end{equation}
The are linear in the chemical potential and the numerical coefficient is
independent 
of the cutoff. In this sense they obey the expected non-renormalization
theorem.

\subsection{Inclusion of metric perturbations}

In this section we compute the flow equations for the Green's functions
associated with generic response. The method can be described  as follows: we
need to consider two equations. One is the constitutive equation

\begin{eqnarray}\label{consdef}  \left<\mathcal{O}_j\right> = \sum_i^N G^i_j
\phi_i \end{eqnarray}

 and the other one is the covariant holographic definition of the one-point
functions, evaluated on some perturbed state. Generically, these would be a
functional of the perturbations and its derivatives (the dot means $d/dr$. We
will be using both notations indistinctly). 

\begin{eqnarray} \label{holodef} \left<\mathcal{O}_j\right> =- \sum_i^N
(\mathcal{F}^i_j \phi_i + \mathcal{H}^i_j \dot \phi_i)\end{eqnarray}

Taking the r-derivative in both equations, we can force them to be equal.
Observe that, from (\ref{holodef}), we expect terms containing $ \mathcal{H}^i_j
\ddot \phi_i$. After using the equations of motion, we will be left with some
expression involving only $\phi$ and $\dot \phi$. Then, by equating
(\ref{consdef}) and (\ref{holodef}), it is possible to find a formula for $\dot
\phi_j =\sum_i^N K^i_j \phi_i $ so that eventually we are able to write the
r-derivative of (\ref{holodef}) as an expansion in the perturbations only.\\
On the other hand, differentiating (\ref{consdef}) and using again $\dot \phi_j
=\sum_i^N K^i_j \phi_i $, we are lead to an expression in terms of $G^i_j$,
$\dot G^i_j$ and $\phi_i$.

Imposing that the r-derivative of (\ref{holodef}) and that of  (\ref{consdef})
are identical, we finally arrive at 

\begin{eqnarray} 0=\sum_i^N A^i_j \phi_i \end{eqnarray}

where $A^i_j$ is a functional of $G^i_j$ and their first derivatives. Assuming
now that the different perturbations are independent from each other, we get $N$
independent equations

\begin{eqnarray} A^i_j =0\end{eqnarray}

which are nothing but differential equations for $G^i_j$. Remarkable enough, the
flow equations for the retarded correlators are of first order in r-derivatives.

\subsubsection{Application to the anomalous conductivities}

In what follows we will derive the flow equations in the presence of a pure
gauge Chern-Simons term (no gravitational anomaly) by using the procedure
detailed in the previous section. The model reads

\begin{eqnarray}\label{AcCSf}   S=  S_{EH}+ S_{GH} + \frac{1}{16\pi
G}\int_{r<\Lambda}
\sqrt{-g}\left(-\frac{1}{4} F_{MN}F^{MN} +  \frac{\kappa}{3} \epsilon^{MNPQR}A_M
F_{NP}F_{QR}\right) \end{eqnarray} 

where $S_{EH}$ denotes the Einstein-Hilbert action with negative cosmological
constant
and $S_{GH}$ is the Gibbons-Hawking term on the boundary $r=\Lambda$.
The Chern-Simons coupling is here related to the anomaly for a single chiral
fermion by $\kappa = -G/(2\pi)$. 

Since we need now the precise equations of motion for the metric fluctuations
we will specialize the analysis to a Reissner-Nordstrom AdS
Black Brane

\begin{eqnarray} \label{background} ds^2 = \frac{r^2}{L^2} \left(-f(r)dt^2+d\vec
x^2\right) + \frac{L^2}{r^2 f(r)}dr^2 \\
A^{(0)}=\phi(r)dt=-\frac{\mu r_H^2}{r^2} \end{eqnarray}

The integration constant in the gauge field is set that it vanishes 
for $r\rightarrow \infty$.
The horizon of the black hole is located at $r=r_H$ and the blackening factor is
$f(r)=1-\frac{ML^2}{r^4} + \frac{Q^2L^2}{r^6}$. The parameters $M, Q$ are
related to the chemical potential at infinity $\mu$ and $r_H$ by $M=
\frac{r_H^4}{L^2}+ \frac{Q^2}{r_H^2}$, $Q=\frac{\mu r_H^2}{\sqrt{3}}$. Finally,
the Hawking temperature is given by

\begin{eqnarray} \label{temp} T =\frac{r_H^2}{4 \pi L^2}\dot{f}(r_H) =
\frac{2r_H^2M-3Q^2}{2 \pi r_H^5} \end{eqnarray}

In what follows, we consider perturbations of momentum $k$ in the $y$-direction
at zero frequency. It is only necessary to turn on the shear sector, that is,
the perturbations are written as $a_{\alpha}$, $h^{\alpha}_t$, where
$\alpha=x,z$ \footnote{At zero frequency the fields $h^{\alpha}_y$ decouple from
the system and thus will not be considered (see \cite{[7]}).}. It is more
convenient to work with the coordinate $u=\frac{r^2_H}{r^2}$ instead of $r$.\\

The equations of motion for the perturbations derived from (\ref{AcCSf}), when
$\omega=0$ and to $\mathcal{O}(k)$, read

\begin{eqnarray}
\label{eq_As}0&=& B_\alpha'' ( u )+\frac { f' ( u ) }{f(u)}B_\alpha'( u ) - 
\frac{ h^{\alpha'}_t ( u
 )}{f(u)} +ik\epsilon_{\alpha\beta}\bar \kappa \frac{B_\beta(u)}{f(u)} \,,\\
\label{eq_Hts} 0&=&  h^{\alpha''}_t(u) - \frac{h^{\alpha'}_t(u)}{u} - 3a u
B'_\alpha(u) \end{eqnarray}

where $\bar \kappa= \frac{4 \mu \kappa L^3}{r_H^2}$. \\

The operators that we will be working with have the following form when
evaluated in a perturbed state (for further details see \cite{[5]})

\begin{eqnarray} \label{currentu} \delta J^{\alpha} = \frac{r_H^2}{8 \pi G L^3}
\left(f(u) a'_{\alpha} - \mu h^{\alpha}_t \right)  \\
 \label{emu} \delta t^{\alpha}_t  =\frac{r^4_H f(u)}{8 \pi G L^5 u}
\left(h'^{\alpha}_t- \frac{3}{u} h^{\alpha}_t \right)  \end{eqnarray}

where the prime stands for $d/du$. Differentiating (\ref{currentu}) and 
(\ref{emu}) we are left with

\begin{eqnarray} \label{currentdifu} \left( \delta J^{\alpha}\right)'
=\frac{r^2_H}{8 \pi G L ^3} \left(a''_{\alpha}(u) f(u) + a'_{\alpha}(u) f'(u)-
\mu h'^{\alpha}_t\right)\\
 \label{emdifu} \left( \delta  t^{\alpha}_t \right)'  = \frac{r^4_H f(u)}{8 \pi
G L^5 u} \left(h''^{\alpha}_t+ h'^{\alpha}_t  \left[
\frac{f'(u)}{f(u)}-\frac{4}{u} \right] + h^{\alpha}_t \left[\frac{6}{u^2}-
\frac{3 f'(u)}{uf(u)} \right] \right)  \end{eqnarray}

In order to handle the $\phi_i''$ terms, we evaluate the above expressions
on-shell, yielding

\begin{eqnarray} \label{currentdifuonshell} \left( \delta J^{\alpha}\right)'
=-\frac{r^2_H}{8 \pi G L ^3}\bar \kappa ik \epsilon_{\alpha \beta} a_{\beta} \\
 \label{emdifuonshell} \left( \delta  t^{\alpha}_t \right)'  = \frac{r^4_H
f(u)}{8 \pi G L^5 u} \left(h'^{\alpha}_t \left[\frac{f'(u)}{f(u)}- \frac{3}{u}
\right] + h^{\alpha}_t \left[\frac{6}{u^2}- \frac{3 f'(u)}{uf(u)} \right] +
\frac{3au}{\mu} a'_{\alpha} \right) \end{eqnarray}

Now, observe that, since  $ h'^{\alpha}_t =\frac{8 \pi G L ^5 u}{r^4_H f(u)}
\delta t^{\alpha}_t + \frac{3}{u} h^{\alpha}_t$ and $ a'_{\alpha} =
\left(\frac{8 \pi GL^3}{r^2_H} \delta J^{\alpha} + \mu h^{\alpha}_t\right)
\frac{1}{f(u)}  $, (\ref{emdifuonshell}) turns into 

 \begin{eqnarray} \label{emdifuonshell2} \left( \delta  t^{\alpha}_t \right)'  =
\frac{3 r^4_H f(u)}{8 \pi G L^5 u} \left[\frac{au}{f(u)}- \frac{1}{u^2}\right]
h^{\alpha}_t   + \left[\frac{f'(u)}{f(u)} -\frac{3}{u} \right]\delta
t^{\alpha}_t + \frac{3 r^2_H a}{L^2 \mu} \delta J^{\alpha}   \end{eqnarray}

Plugging the constitutive relations ($\epsilon^{xz} \equiv 1$)

\begin{eqnarray} \label{constj} \delta J^{\alpha}_{const} =
G^{xx}\delta^{{\alpha}\beta} a_{\beta} + G^{xz} \epsilon^{{\alpha}\beta}
a_{\beta}  + P^{xt}\delta^{{\alpha}\beta} h^t_{\beta} +P^{zt}
\epsilon^{{\alpha}\beta} h^t_{\beta}    \\
\label{constt} \delta t^{\alpha}_{t \ const} =
G_{\epsilon}^{xx}\delta^{{\alpha}\beta} a_{\beta} + G_{\epsilon}^{xz}
\epsilon^{{\alpha}\beta} a_{\beta}  + P_{\epsilon}^{xt}\delta^{{\alpha}\beta}
h^t_{\beta} +P_{\epsilon}^{zt} \epsilon^{{\alpha}\beta} h^t_{\beta} 
\end{eqnarray}

into (\ref{emdifuonshell2}), the remaining equation for $\left( \delta 
t^{\alpha}_t \right)'$ involves only $\phi_i$ and $G^i_j$.  \\

On the other hand, we can take the u-derivative of (\ref{constj})-(\ref{constt})
explicitly and then make use of (\ref{currentu})-(\ref{emu}) to end up having an
equation in terms of $\phi_i$, $G^i_j$ and $G'^i_j$.\\

Finally, imposing $\left(\delta J^{\alpha}_{const}\right)' = \left(\delta
J^{\alpha}\right)'$ and $\left( \delta t^{\alpha}_{t \ const}\right)'= \left(
\delta t^{\alpha}_{t}\right)' $ and assuming that the perturbations $\phi_i$ are
independent from each other, we find

\begin{eqnarray} \label{flow1} G'^{xx} + \frac{8\pi GL^3}{f(u) r_H^2} \left(
\left(G^{xx}\right)^2 -\left( G^{xz}\right)^2 \right) - \frac{ 8 \pi GL^5 u}{
r_H^4 f^2(u)} \left(P^{xt}G^{xx}_{\epsilon} - P^{zt}G^{xz}_{\epsilon} \right) =0
\\
  \label{flow2} G'^{xz} + \frac{16\pi G L^3}{f(u) r_H^2}G^{xx}G^{xz} - \frac{8
\pi G L^5 u}{ r_H^4 f^2(u)} \left(P^{xt}G^{xz}_{\epsilon} +
P^{zt}G^{xx}_{\epsilon} \right) =-\frac{r_H^2}{8 \pi GL^3} \bar \kappa ik \\
\nonumber \label{flow3} P'^{xt} + G^{xx} \left( -\mu + \frac{8 \pi G L^3}{f(u)
r_H^2}P^{xt} \right) -\left( \frac{8 \pi G L^3}{f(u) r_H^2} G^{xz} - \frac{8\pi
GL^5 u}{r_H^4 f^2(u)}P^{zt}_{\epsilon} \right) P^{zt} + \\ +P^{xt}
\left(-\frac{8\pi GL^5u}{r_H^4 f^2(u)}P^{xt}_{\epsilon} -\frac{f'(u)}{f(u)} +
\frac{3}{u}\right)  =0  \\
\nonumber  \label{flow4} P'^{zt}+ G^{xz} \left( -\mu+ \frac{8 \pi G L^3}{f(u)
r_H^2}P^{xt} \right) -  \frac{8 \pi G L^5 u}{f^2(u) r_H^4}
P^{xt}P^{zt}_{\epsilon}+\\ +P^{zt} \left(-\frac{8 \pi G L^5 u}{f^2(u)
r_H^4}P^{xt}_{\epsilon} + \frac{3}{u} - \frac{f'(u)}{f(u)} + \frac{8\pi G
L^3}{f(u) r_H^2}G^{xx}\right) =0 \end{eqnarray}

\begin{eqnarray} \label{flow5} \nonumber G'^{xx}_{\epsilon} + \frac{8\pi
GL^3}{f(u) r_H^2} \left( G^{xx}_{\epsilon}G^{xx} -G^{xz}_{\epsilon}G^{xz}
\right) - \frac{ 8 \pi GL^5 u}{ r_H^4 f^2(u)} P^{xt}_{\epsilon}
G^{xx}_{\epsilon} + \frac{ 8 \pi GL^5 u}{ r_H^4
f^2(u)}P^{zt}_{\epsilon}G^{xz}_{\epsilon} =\\= -G^{xx}_{\epsilon}
\left(\frac{3}{u} - \frac{f'(u)}{f(u)}\right) + \mu G^{xx} \\
  \label{flow6} \nonumber  G'^{xz}_{\epsilon} + \frac{8\pi G L^3}{f(u)
r_H^2}\left(G^{xx}_{\epsilon}G^{xz} +G^{xz}_{\epsilon}G^{xx}  \right) - \frac{8
\pi G L^5 u}{ r_H^4 f^2(u)} \left(P^{xt}_{\epsilon} G^{xz}_{\epsilon} +
P^{zt}_{\epsilon}G^{xx}_{\epsilon} \right) =\\ = -G^{xz}_{\epsilon}
\left(\frac{3}{u}-\frac{f'(u)}{f(u)}\right) + \mu G^{xz} \\
 \label{flow7} \nonumber P'^{xt}_{\epsilon} + G^{xx}_{\epsilon} \left(-\mu +
\frac{8 \pi G L^3}{f(u) r_H^2}P^{xt} \right) - \frac{8 \pi G L^3}{f(u) r_H^2}
G^{xz}_{\epsilon}P^{zt}+ P^{xt}_{\epsilon} \left( -\frac{8\pi GL^5 u}{r_H^4
f^2(u)}P^{xt}_{\epsilon}  +\frac{3}{u} - \frac{f'(u)}{f(u)} \right)  + \\+
\frac{8\pi GL^5 u}{r_H^4 f^2(u)}\left(P^{zt}_{\epsilon} \right)^2 =
-P^{xt}_{\epsilon}\left(\frac{3}{u} - \frac{f'(u)}{f(u)}\right)+ \mu
P^{xt}-\frac{3 r_H^4}{8 \pi GL^5u}f(u)\left(au-\frac{f(u)}{u^2}\right)   \\
 \label{flow8}\nonumber P'^{zt}_{\epsilon}+ G^{xz}_{\epsilon} \left(  \frac{8
\pi G L^3}{f(u) r_H^2}P^{xt} -\mu \right) +\frac{8 \pi G L^3 }{f(u) r_H^2}
G^{xx}_{\epsilon}P^{zt}-  \frac{8 \pi G L^5 u}{f^2(u) r_H^4}
P^{xt}_{\epsilon}P^{zt}_{\epsilon}+\\ +P^{zt}_{\epsilon} \left(-\frac{8 \pi G
L^5
u}{f^2(u) r_H^4}P^{xt}_{\epsilon} + \frac{3}{u} - \frac{f'(u)}{f(u)} \right)
=-P^{zt}_{\epsilon}\left(\frac{3}{u} -\frac{f'(u)}{f(u)}\right) + \mu P^{zt}
\end{eqnarray}

By directly studying
the structure of the solutions to (\ref{eq_As})-(\ref{eq_Hts}), it can be
realized that $G^{xx}= P^{xt}=G^{xx}_{\epsilon}=0$ for $\omega=0$ and to 
first order in $k$.
Furthermore, all the anomalous correlators are of order $k$ or higher. A more
detailed study of (\ref{flow1})-(\ref{flow8}) is left for section
\ref{sec:compatibility}.

\subsection{Flow of the transport coefficients as two point functions}

As suggested in Section 2, we could have determined the flow by simply
considering the system to be restricted to live between the horizon and a cutoff
surface placed at $\Lambda$. It is hence expected that the transport
coefficients at the boundary can be computed by finding the
corresponding 2-point functions. The boundary value of the perturbations, whose
bulk-to-boundary propagator is normalized at the cutoff, work as the sources for
the different operators of the dual theory. \\

Henceforth, the perturbations will be rearranged in a vector $\Phi (u,
x^{\mu})$. It is more convenient to use the Fourier transformed quantity

\begin{eqnarray} \Phi (u, x^{\mu}) = \int \frac{d^d k}{(2 \pi)^d} \Phi_k^I (u)
e^{-i\omega t + i \vec k \vec x}\end{eqnarray}

The explicit expression for $\Phi_k (u)$ is

\begin{eqnarray} \Phi_k^\top (u) = \left(B_x(u), h^x_t (u),  B_z (u), h^z_t(u)
\right) \end{eqnarray} 

being $B_{\alpha}= a_{\alpha}/\mu$. To proceed, one can follow \cite{[9]} and
assume the general form of a boundary action

\begin{eqnarray} \label{deltact} \delta S^{(2)}=\int_{r=\Lambda}
\frac{d^dk}{(2\pi)^d}\left[\Phi^I_{-k}\mathcal{A}_{IJ}\Phi'^J_{k}
+\Phi^I_{-k}\mathcal{B}_{IJ}\Phi^J_{k}  \right] \end{eqnarray}

In order to get the solution of the system (\ref{eq_As})-(\ref{eq_Hts}) to first
order in momentum we expand the fields in the (dimensionless) quantity
$p=\frac{k}{4 \pi T}$. Hence

\begin{eqnarray} h^{\alpha}_t(u) =h^{(0),\alpha}_t(u) + p h^{(1),\alpha}_t(u)\\
B_{\alpha} (u)= B^{(0)}_{\alpha} (u)+ pB^{(1)}_{\alpha}(u) \end{eqnarray} 

The system can be solved perturbatively . 
To calculate the retarded correlators at $r=\Lambda$ (or, equivalently, at
$u=u_c \equiv r_H^2/\Lambda^2$) we only need to solve the equations for the
perturbations with infalling boundary conditions, on the one hand, and boundary
conditions $\Phi^I_k (u_c) = \phi^I_k$ on the other \cite{[5]}. This procedure
should give us the desired Green's functions, after taking the variation of
(\ref{deltact}) with respect to the fields at $u=u_c$ (which act as sources for
their corresponding operators). Recall that, as explained in Section 2, the
bulk-to-boundary propagator must be normalized at $r=\Lambda$, that is, if we
have 

\begin{eqnarray} \Phi^I_{k}(u) = F^I_J(k,u) \phi^J_k \end{eqnarray}

then $ F^I_J(k,u_c)=1$.  Notice that the relation between the boundary value at
$u=u_c$ and that at $u=0$ is simply $\phi^{I\ (u_c)}_{k} =F^I_J(k,u_c)  \phi^{J\
(0)}_{k} $, so that the solution is preserved by these manipulations, as pointed
out by \cite{[4]} and \cite{[8]}. The retarded two-point functions, from which
we are able to read directly the transport coefficients, then have the form

 \begin{eqnarray} \label{defcorrel} G_{IJ} (k, u_c) = -2 \lim_{u \rightarrow
u_c} \left(\mathcal{A}_{IM}\left(F^M_J(k,u)\right)'+ \mathcal{B}_{IJ} \right)
\end{eqnarray}

Where the $\mathcal{A}_{IJ}$ and $\mathcal{B}_{IJ}$ matrices are \cite{[5]}

\begin{eqnarray} \mathcal{A} = \frac{r^4_H}{16 \pi G L^5} Diag \left(-3af(u),
\frac{1}{u}, -3af(u), \frac{1}{u} \right)\end{eqnarray}

\begin{eqnarray} B_{AdS +\partial}=\frac{r^4_H}{16 \pi G L^5}\begin{bmatrix}
0 & 3a & 0 & 0 \\
0 & -\frac{3}{u^2} & 0 & 0 \\
0 & 0 & 0 & 3a \\
0 & 0 & 0 & -\frac{3}{u^2}
\end{bmatrix}\end{eqnarray}

Using again the  the effective chemical potential
 
\begin{eqnarray} \label{mu} \mu(\Lambda) =\mu\left(1-
\frac{r_H^2}{\Lambda^2}\right)\,,
\end{eqnarray}  

the result for the anomalous correlators is 

\begin{eqnarray} 
\label{CMEff} \left<\delta J^x \delta J^z  \right> =
\frac{i\mu\kappa k}{2\pi G}\left( 1- \frac{r_H^2}{\Lambda^2}\right)
&= &\frac{-i k \mu(\Lambda)}{4\pi^2}\\
\label{CVEff}
\left<\delta J^x \delta t^z_t  \right> = \left<
\delta t^x_t  \delta J^z \right> =-\frac{i \kappa \mu^2 k}{4\pi G}\left( 1-
\frac{r_H^2}{\Lambda^2} \right)^2
&=&\frac{i k \mu(\Lambda)^2}{8\pi^2}
\\ \label{CVEEff}
 \left<\delta t^x_t \delta t^z_t  \right> = \frac{i \kappa \mu^3
k}{6\pi G}\left( 1- \frac{r_H^2}{\Lambda^2} \right)^3 &= &\frac{-ik
\mu(\Lambda)^3}{12\pi^2}
\end{eqnarray}

Since $\lim_{\Lambda\rightarrow\infty}\mu(\Lambda) = \mu$, these correlators
coincide essentially with
the ones derived in \cite{[5]}\footnote{The minus sign found in (\ref{CVEff})
with respect to
the result of \cite{[5]} is due to the fact that in this reference the
correlator that is studied is $\left<\delta J^a \delta t_b^{t}  \right>$, that
differs from $\left<\delta J^a \delta t^{b}_t  \right>$ by a factor of ($b$ represents a spatial index) 
$g_{tt}g^{bb} = -f(u) \rightarrow -1$ at infinity.}.\\

\subsubsection{Compatibility with the flow equations}
\label{sec:compatibility}
The system of first order differential equations (\ref{flow1})-(\ref{flow8})
must be compatible with the result  (\ref{CMEff})-(\ref{CVEEff}) encountered in
the previous section. In order to check that it is so, the dissipative
correlators play an important role. In the case $\omega=0$ and to
$\mathcal{O}(k)$, they read\footnote{The limit
$P^{xt}_{\epsilon}(u=0)$ is not well defined because we have not included the
corresponding counterterms in (\ref{currentu}),(\ref{emu}). The reason is that
they do not affect the anomalous correlators.}

\begin{eqnarray} \label{diss1} G^{xx} = P^{xt} = G^{xx}_{\epsilon} = 0 \\
\label{diss2} P^{xt}_{\epsilon} = -\frac{r_H^4}{8 \pi G L^5u}f^2(u)
\left(\frac{f'(u)}{f(u)} - \frac{3}{u} \right) \end{eqnarray}

This solution implies that $G^{xx}$ and $P^{xt} = G^{xx}_{\epsilon}$ are of
order $\omega$ or higher, whereas $P^{xt}_{\epsilon}$ contains a part which is
of order $\mathcal{O}(k^0, \omega^0)$ (contact term). The remaining system,
after substituting (\ref{diss1}), (\ref{diss2}) and assuming that all the
anomalous correlators are at least of $\mathcal{O}(k)$, turns out to be (up to
order $k$)

\begin{center}
\begin{eqnarray} \label{flow9} G'^{xx} =0 \\
  \label{flow10} G'^{xz}=-\frac{r_H^2}{8 \pi GL^3} \bar \kappa ik \\
\label{flow11} P'^{xt}   =0  \\
 \label{flow12} P'^{zt} -\mu G^{xz} =0 \\
 \label{flow13}G'^{xx}_{\epsilon} =0 \\
 \label{flow14}  G'^{xz}_{\epsilon} = \mu G^{xz} \\
\label{flow15} P'^{xt}_{\epsilon}  = -P^{xt}_{\epsilon}\left(\frac{3}{u} -
\frac{f'(u)}{f(u)}\right)-\frac{3 r_H^4}{8 \pi
GL^5u}f(u)\left(au-\frac{f(u)}{u^2}\right)   \\
 \label{flow16}  P'^{zt}_{\epsilon}- \mu G^{xz}_{\epsilon}  = \mu P^{zt}
\end{eqnarray}
\end{center}

Equation (\ref{flow15}) is in agreement with (\ref{diss2}). In the end, the
2-point functions associated with dissipative transport coefficients decouple
completely. Regarding the anomalous correlators, the above system of equations
can be integrated easily, leading to

\begin{eqnarray} \label{corr1} G^{xz}= \frac{r_H^2}{8 \pi GL^3} \bar \kappa ik
\left(1-u_c\right) \\
\label{corr2} P^{zt} = G^{xz}_{\epsilon} =  -\mu \frac{r_H^2}{16 \pi GL^3} \bar
\kappa ik \left(1-u_c\right)^2 \\
\label{corr3} P^{xt}_{\epsilon} =  \mu^2 \frac{r_H^2}{24 \pi GL^3} \bar \kappa
ik \left(1-u_c\right)^3\end{eqnarray}

which is the same as (\ref{CMEff})-(\ref{CVEEff}). The role played by the
Chern-Simons term in (\ref{AcCSf}) is crucial to ensure that $G^{xz}$ presents a
flow, for in its absence all the anomalous 2-point functions identically vanish.

\section{Gravitational Anomaly}

The study of the effect of the Gravitational Anomaly on the definition of the
holographic operators is a non-trivial task, for the term $A \wedge R \wedge R$
has not a well defined Dirichlet problem. This makes, strictly speaking, not
possible to define generic operators. In \cite{[7]}, the problem was
circumvented by arguing that any possible contribution vanishes asymptotically.
However, now we are interested in the value of the transport coefficients at
finite cutoff $\Lambda$, and therefore it is necessary to face this issue.

\subsection{The Model} 

The four dimensional axial gravitational anomaly is induced holographically by a
Chern-Simons term of the form \cite{[7]}

\begin{eqnarray} \label{31} S_{\mathcal{A}CS} = \frac{\lambda}{16 \pi G} \int
d^5x \sqrt{-g} \epsilon^{MNPQR} A_{M}R^{A}_{\ BNP}R^B_{\ AQR} \end{eqnarray}

This action contributes to the boundary axial current as expected for a mixed 
anomaly. The complete action reads 

\begin{eqnarray} \label{32} S= \frac{1}{16 \pi G} \int d^5 x \sqrt{-g} \left[ R
+ 2 \Lambda_{c} - \frac{1}{4}F_{MN} F^{MN}\right] + S_{\mathcal{A}CS} +
S_{\mathcal{A}EM}+  S_{\partial} +S_{CSK} \end{eqnarray}

 Where
\begin{align}
 S_{\mathcal{A}EM} &= \frac{\kappa}{48 \pi G}\int d^5x \sqrt{-g}
\epsilon^{MNPQR} A_M F_{NP}F_{QR} \,,\\
 S_{\partial} &= -\frac{1}{8 \pi G} \int_{\partial} \sqrt{-h}K\,,\\
 S_{CSK} &= - \frac{\lambda}{2 \pi G} \int_{\partial \mathcal{M}} d^4 x
\sqrt{-h} n_M \epsilon^{MNPQR} A_N K_{PL} D_Q K^{L}_R \,.
\end{align}

Adding $S_{CSK}$ ensures that the anomalous Ward identity for gauge
transformations
depends only on the intrinsic curvature tensor on the boundary at $r=\Lambda$
\cite{[7]}.

Indeed, the covariant current turns out to be

\begin{eqnarray} 16 \pi G J^A = n_B \left[F^{AB} - 8 \epsilon^{BACDE}  \lambda
K_{CF} D_D K^E_F  \right]_{ r=\Lambda} 
\end{eqnarray}
 with a purely four dimensional divergence that on shell evaluates to

\begin{eqnarray} \label{38} D_{\mu} J^{\mu} = -\frac{1}{16 \pi G}
\epsilon^{opqr} \left[ \frac{\kappa}{3} F_{op} F_{qr} + \lambda
R^{a}_{(4)bop}R^{b}_{(4)aqr}\right]_{r =\Lambda} \end{eqnarray}

where $\epsilon^{opqr} \equiv \epsilon^{\textbf{n}opqr} $ is the four
dimensional epsilon tensor.\\

The bulk equations of motion are

\begin{eqnarray}\label{eqgrav}
 G_{MN} - \Lambda_c g_{MN} &=& \frac 1 2 F_{ML} F_N\,^L - \frac 1 8 F^2 g_{MN} +
2 \lambda \epsilon_{LPQR(M} \nabla_B\left( F^{PL} R^B\,_{N)}\,^{QR} \right) \,,
\label{eq:Gbulk}\\\label{eqgauge}
\nabla_NF^{NM} &=& - \epsilon^{MNPQR} \left( \kappa F_{NP} F_{QR} + \lambda 
R^A\,_{BNP} R^B\,_{AQR}\right) \,,  \label{eq:Abulk}
\end{eqnarray}

\subsection{Contribution of the Gravitational Anomaly} 

If we vary $S_{\mathcal{A}CS}$, we are left with a term which spoils the
variational problem

\begin{eqnarray}\label{rem1} \frac{\lambda}{2 \pi G} \int_{\partial} \sqrt{-h}
\epsilon^{mlqr}  A_m D_r  K^v_q \delta K_{lv}\end{eqnarray}

If we looked for a suitable counterterm to render the Dirichlet problem
well-posed,  we would end up finding $S_{CS\mathcal{K}}$. Indeed, this boundary
contribution was firstly conceived as an analogue to the Gibbons-Hawking-York
term. However, after varying $S_{\mathcal{A}CS} +S_{CS\mathcal{K}} $ one
realizes that the result

\begin{eqnarray}\label{rem2} -\frac{\lambda}{2 \pi G} \int_{\partial} \sqrt{-h}
\epsilon^{mlqr} D_r A_m \delta K^v_q K_{lv}\end{eqnarray}

is still problematic. Even worse, (\ref{rem2}) can not be canceled easily, for,
for instance, the ansatz

\begin{eqnarray}  \frac{\lambda}{2 \pi G} \int_{\partial} \sqrt{-h}
\epsilon^{mlqr} D_r A_m K^v_q K_{lv}\end{eqnarray} 

is automatically zero. Thus in principle, there is not a straightforward way of
having a well defined variational problem for this system. \\

On the other hand, as aforementioned, we need $S_{CS\mathcal{K}}$ to have a four
dimensional anomalous Ward identity at the boundary, so we will keep it. A
hypothetical generic counterterm (if it exists) capable of solving all the
problems, would probably ruin (\ref{38}) and therefore, by physical means,
should not be considered.

Even though the variational problem is not well-posed, we will still be able to
derive the equations of motion by means of the analogue of the Euler-Lagrange
equations for higher-derivative theories. The difficulty therefore reduces to
the question \emph{How to treat (\ref{rem2}) holographically?} Note that in
\cite{[9]} it was implicitly assumed that the Dirichlet problem is correctly
defined, so we should go a little bit further in this case.

Specializing for the shear sector, which is the one that interests us, and at
second order in perturbations, (\ref{rem2}) reads

\begin{eqnarray} \label{tom} -\frac{\lambda}{2 \pi G} \int_{\partial} \sqrt{-h}
\epsilon^{mlqr} D_r \delta A_m \delta K^v_q K_{lv} \end{eqnarray}

Other possible terms would vanish in the background (\ref{background}). 
The strategy would be the following: Since
(\ref{tom}) does not affect two point functions involving only energy-momentum
tensors or only currents, we know how to compute the correlators $\left<T^x_t
T^z_t \right>$ and $\left<J^x J^z \right>$.  (\ref{tom}) only plays a role when
calculating  $\left<T^x_t J^z \right>$, $\left<J^x T^z_t \right>$, and hence
those are the ones for which the discussion of \cite{[9]} does not apply.\\
Following the method detailed in Section 3, it turns out that, taking only into
account the gravitational anomaly

\begin{eqnarray}\label{aer} \left<J^x J^z \right> = 0\\
\label{air} \left<T^x_t T^z_t \right> =-ik \frac{\mu (1-u_c)T^2}{12}
\end{eqnarray}
 
(note that we have directly substituted the value of $\lambda$ for a single
left-handed fermion $\lambda/G= -\frac{1}{48\pi}$). The above results point
again towards an effective $\mu(1-u_c)$. Therefore, by physical grounds, we
expect the appearance of an effective temperature also.  Note that the flows of
the effective quantities must be consistent in the sense that they must be the
same, no matter what correlator we are focusing on. Equations
(\ref{aer})-(\ref{air}) hint at the existence of an effective temperature for
the system; this temperature does not flow with the cutoff scale, being always
identical to the Hawking temperature. This conclusion is in agreement with the
asymptotic values of \cite{[7]}.\\

So we resolve that (\ref{rem2}) must be treated in such a way that $\left<T^x_t
J^z \right>$, $\left<J^x T^z_t \right>$, at finite cutoff, are consistent with a
non-flowing temperature. \\
It turns out that the method to achieve it is precisely the one that one would
anticipate by general considerations: Taking advantage of the fact that the
equations of motion for the shear sector

\begin{eqnarray}
\label{htx1} \nonumber 0&=&  h^{\alpha''}_t(u) - \frac{h^{\alpha'}_t(u)}{u} 
- 3a u B'_\alpha(u) +i\bar\lambda k \epsilon_{\alpha\beta}\left[\left(24a
u^3-6(1-f(u))\right)\frac {B_\beta(u)}{u}\right.\\
    &&\left.+(9a u^3-6(1-f(u)))B'_\beta(u)+2 u (uh^{\beta'}_t ( u ) )' \right]
\,, \\ 
\nonumber\label{bx1} 0&=& B_\alpha'' ( u )+\frac { f' ( u ) }{f(u)}B_\alpha'( u
)-  \frac{ h^{\alpha'}_t ( u
 )}{f(u)} \\
 &&+ik\epsilon_{\alpha\beta}\left(\ \frac{3}{u f(u)} \bar\lambda \left(
\frac{2}{a} (f(u)-1)+ 3u^3 \right) h_t^{\beta'}(u)+\bar \kappa
\frac{B_\beta(u)}{f(u)} \right) \, ,
\end{eqnarray}

 happen to be of second order in
derivatives (where $\bar \lambda= \frac{4
\mu \lambda L}{r_H^2}$), we can solve completely the
evolution as we did in Section 3.3 (imposing in-falling B.C. at the Horizon and
Dirichlet B.C. at the boundary). Once the solutions are known (see the
appendix), (\ref{rem2}) will
in general give a well determined surface contribution (when evaluated on-shell)
that must be taken into account to calculate  $\left<T^x_t J^z \right>$,
$\left<J^x T^z_t \right>$. The result so obtained presents no flow in the
temperature part.\\

To be more concise, the boundary term (\ref{rem2}) to be considered has the
following form

\begin{eqnarray} -\frac{ik\lambda r_H^2 \epsilon_{\alpha \beta} }{2 \pi G L^4}
\int_{\partial} u f'(u) a_{\beta}(k) h'^{\alpha}_t(-k)   \end{eqnarray}  

whose contribution, up to first order in $k$, is summarized

\begin{eqnarray} \label{contprev} -\frac{ik\lambda r_H^2 \epsilon_{\alpha \beta}
}{2 \pi G L^4} \int_{\partial} u \frac{f'^2(u)}{f(u_c)} a^{(0)}_{\beta}(k)
\tilde H^{(0)}_{\beta}(-k) \end{eqnarray}

(Notice the factor $\sim \frac{1}{f(\Lambda)}$ introduced to normalize the
perturbation (see the appendix)). So the effect of (\ref{rem2}) on
the Green's functions can be reformulated as a modification, prescribed by
(\ref{contprev}), of the $\mathcal{B}_{IJ}$ matrix.\\

 Even though (\ref{contprev}) only affects the correlator $\left<T^{\alpha}_t
J^{\beta}\right>$, $S_{\mathcal{ACS}} +S_{\mathcal{CSK}}$ 
induces automatically a non-vanishing value for the components
$\mathcal{A}_{14}=\mathcal{A}^*_{32}$ of the matrix $\mathcal{A}$. 
These contributions, which are perfectly treatable within the framework of
\cite{[9]}, give rise to a correction of  $\left<J^{\alpha}T^{\beta}_t \right>$
which is precisely of the same form of the one implemented by (\ref{contprev}).
As will be mentioned below, this turns out to be sufficient for the consistency
condition (\ref{condgreen}) to hold.\\

The final form of the matrices $\mathcal{A}_{IJ}$ and $\mathcal{B}_{IJ}$ after
implementing the shift driven by the Gravitational Anomaly is given by

\begin{eqnarray} \mathcal{A} = \frac{r^4_H}{16 \pi G L^5} \begin{bmatrix}
-3af(u) & 0 & 0 & -\frac{4 i\lambda k L}{r_H^2} u f'(u) \\
0 & \frac{1}{u} & 0 & \frac{i 8 \lambda k L \mu}{r_H^2} u \\
0 & \frac{4 i\lambda k L}{r_H^2} u f'(u) & -3af(u) & 0 \\
0 & -\frac{i 8 \lambda k L \mu}{r_H^2} u & 0 & \frac{1}{u} 
\end{bmatrix}\end{eqnarray} 

\begin{eqnarray} B_{AdS +\partial}=\frac{r^4_H}{16 \pi G L^5}\begin{bmatrix}
0 & 3a & 0 & 0 \\
0 & -\frac{3}{u^2} & 4 i \lambda k L \frac{9 a u^3 - 6 (1 - f(u)) }{u r_H^2}  &
0 \\
0 & 0 & 0 & 3a \\
-4 i \lambda k L \frac{9 a u^3 - 6 (1 - f(u)) }{u r_H^2} & 0 & 0 &
-\frac{3}{u^2}
\end{bmatrix}\end{eqnarray}

\begin{eqnarray} B_{\partial CS}=\frac{r^4_H}{16 \pi G L^5}\begin{bmatrix}
0 & 0 & 0 & 0 \\
0 & 0 & -\frac{4 i \lambda k L}{r_H^2} u \frac{f'(u)^2}{f(u_c)}  & 0 \\
0 & 0 & 0 & 0 \\
\frac{4 i \lambda k L}{r_H^2} u \frac{f'(u)^2}{f(u_c)} & 0 & 0 & 0
\end{bmatrix}\end{eqnarray}

The resulting anomalous 2-point functions are

\begin{align} 
\label{CMElam} \left<J^x J^z \right> &=  {\frac {\,ik\kappa\,\mu  \left( 1-u_c
\right) }{2G \pi }} = - ik \frac{\mu(\Lambda)}{4 \pi^2}\\
\label{CVElam} \left<J^x T^z_t \right> &=  -{\frac {\,ik\kappa\, \left(1-u_c
\right) ^{2}{\mu}^{2}}{4G\pi}}-{\frac {\,i k \lambda \left( -2+a \right)
^{2}{{\it r^2_H}}}{2G{L}^{4}\pi }} =ik\left( \frac{\mu^2 (1-u_c)^2}{8\pi^2} +  
\frac{T^2}{24}\right)\\
\label{CVElam2} \left<T^x_t J^z \right> &= - {\frac {\,ik\kappa\, \left(1-u_c
\right) ^{2}{\mu}^{2}}{4 G\pi}}-{\frac {\,i k \lambda \left( -2+a \right)
^{2}{{\it r^2_H}}}{2G{L}^{4}\pi }}= ik\left( \frac{\mu^2 (1-u_c)^2}{8\pi^2} +  
\frac{T^2}{24}\right) \\ \label{CVEElam}
\left<T^x_t T^z_t \right>&= {\frac {\,ik\kappa\, \left( 1-u_c \right)
^{3}{\mu}^{3}}{6G\pi }}+\, \left( 1-u_c \right) \mu {\frac {ik \lambda \left(
-2+a \right) ^{2}{{\it r^2_H}}}{G{L}^{4}\pi }} =-ik \left(\frac{\mu^3
(1-u_c)^3}{12\pi^2}+\frac{\mu (\Lambda)T^2}{12} \right) 
 \end{align} 

Observe that it is straightforward to verify that equations
(\ref{CMElam})-(\ref{CVEElam}) are compatible with the asymptotic value computed
in \cite{[7]}. Notice also that the temperature part remains constant as we move
the boundary. The flow of the different correlators is consistent with respect
to each other and the hypothesis of an effective chemical potential
$\mu(\Lambda) =  \mu\left(1-\frac{r_H^2}{\Lambda^2}\right) \equiv \mu(1-u_c)$ is
reinforced by the results extracted from the terms proportional to $\lambda$.\\

\section{Discussion and Conclusion}

We have studied the holographic cutoff flow of the anomalous transport
coefficients. This has been done by defining a bottom up model that implements
both the axial and the mixed axial-gravitational anomalies. The flow has been
studied by analyzing the dependence of the anomalous Green's functions on the
radial position, $\Lambda$, of the boundary. We have presented several
prescriptions to compute such flow and finally obtained it by adapting the
method implemented in \cite{[7]}, \cite{[5]} for
the case $\Lambda \rightarrow \infty$.

It is a remarkable fact that the chiral magnetic conductivity suffers from a
flow even in the non-backreacted case. In fact, this could have been anticipated
by noticing that regularity at the horizon imposes that in the deep IR the
constitutive relations are only compatible with an electric conductivity
(\cite{[4]}), so that if a system exhibits a chiral magnetic conductivity in the
UV it must be due to a non-trivial flow.

When considering the gravitational anomaly, a Dirichlet boundary condition is
not enough anymore to define the variational problem properly. A generic
definition of suitable operators, if any, therefore requires further discussion
in this case. In this paper we have simply focused on computing $2$-point
functions, without discussing general definitions of the
corresponding operators. The term which spoils the variational principle has
been dealt with by considering its effect on the on-shell action. This
procedure, which can be seen to be the most natural one by using physical
arguments, yields 2-point functions that are consistent and whose flows do not
get in contradiction with the result found in the absence of gravitational
anomaly. Moreover, in the spirit of \cite{[9]}, that the  matrix of correlators 
$\mathcal{G}_{IJ}$ obeys

\begin{eqnarray}  \label{condgreen} \frac{d}{du} \left( \mathcal{G} - 
\mathcal{G}^{\dagger} \right) =0 \,,\end{eqnarray}

represents a non-trivial consistency check. 

The result (\ref{CMElam})-(\ref{CVEElam}) shows that the temperature remains
constant (Hawking temperature) whereas the chemical potential presents a flow
that is easily interpretable in terms of the energy necessary to bring a unit of
charge from the horizon to the boundary. Observe, however, that all the
correlators are written for a metric with $g_{tt} \sim -r^2 f(r)$, and hence
there is an implicit redshift factor between observers living in one
hypersurface placed at $r= \Lambda$ and another one at $r= \Lambda'$.  \\
From the point of view of the boundary theory, these outcomes indicate that the
pure gauge Chern-Simons term does not affect the boundary operators but
influences the anomalous correlators through the flow equations, forcing them to
have a non-vanishing value at the boundary, whereas the gravitational-gauge
Chern-Simons term happens not to have any impact by means of the evolution
equations, but to induce new covariant contributions, that are first order in
$k$, to the operators, so that the constant $T^2$ part is present at any value
of the $r$-coordinate. 

\section*{Appendix:\\ Solutions at zero frequency and normalized at finite
cutoff
$u_c$}

\subsection*{Case $\lambda=0$}

\begin{align}\nonumber B^{\alpha}(u) =\bar B^{\alpha}+\bar H^{\alpha} (u-u_c
) -\frac{i\bar \kappa k \epsilon_{\alpha \beta}}{2 (1+4 a)^2 (-1+u_c  (-1+a u_c
))} \times \\\nonumber  \times ((1+4 a) (u-u_c ) (\bar H^{\beta}+\bar H^{\beta}
u_c +a (3 \bar B^{\beta} (2+u_c )+\bar H^{\beta} (4-u_c  (2+3 u_c
))))+\\\nonumber +2 \sqrt{1+4 a} (-2+a (-2+3 u)) (\bar B^{\beta}-\bar H^{\beta}
u_c ) (-1+u_c  (-1+a u_c )) (\text{ArcTanh}\left[\frac{-1+2 a u}{\sqrt{1+4
a}}\right]+\\ +\text{ArcTanh}\left[\frac{1-2 a u_c }{\sqrt{1+4
a}}\right]))\end{align}

\begin{align} \nonumber H^{\alpha}_t(u)= -\frac{1}{2 (-1-4 a)^{3/2}
(-1+u_c)(-1+u_c  (-1+a u_c ))^2}(-1+u) \times \\ \nonumber \times (-2 (-1-4
a)^{3/2} \bar H^{\alpha} (-1+u (-1+a u)) (-1+u_c  (-1+a u_c )) +\\ \nonumber +k
\bar \kappa \epsilon_{\alpha \beta} (-i \left(\sqrt{-1-4 a}-i \sqrt{1+4
a}\right) \bar H^{\beta} (1+u) (1+u_c ) +\\\nonumber +a^2 3 \bar B^{\beta}
\left(2 i \sqrt{-1-4 a} u_c ^2+i \sqrt{-1-4 a} u u_c ^2+\sqrt{1+4 a} u^2 (2+u_c
)\right)+ \\\nonumber+ a^2 \bar H^{\beta}  \left(2 i \sqrt{-1-4 a} (2-3 u_c )
u_c ^2+i \sqrt{-1-4 a} u (4-3 u_c ) u_c ^2+\sqrt{1+4 a} u^2 (4-u_c  (2+3 u_c
))\right) -\\\nonumber -3 i a\bar B^{\beta} (2 \sqrt{-1-4 a}-2 i \sqrt{1+4 a}+2
\sqrt{-1-4 a} u_c -i \sqrt{1+4 a} u_c) -\\\nonumber - 3iau \bar
B^{\beta}\left(\sqrt{-1-4 a}-2 i \sqrt{1+4 a}+\sqrt{-1-4 a} u_c -i \sqrt{1+4 a}
u_c \right) +\\ \nonumber + a\bar H^{\beta} (-4 i \sqrt{-1-4 a}-4 \sqrt{1+4
a}+\sqrt{1+4 a} u^2 (1+u_c )) +\\\nonumber +a\bar H^{\beta} u_c  \left(2 i
\sqrt{-1-4 a}+2 \sqrt{1+4 a}+7 i \sqrt{-1-4 a} u_c +3 \sqrt{1+4 a} u_c
\right)+\\\nonumber +ua\bar H^{\beta} (-4i \sqrt{-1-4 a}-4 \sqrt{1+4 a}
)+\\\nonumber + ua\bar H^{\beta} u_c  \left(-i \sqrt{-1-4 a}+2 \sqrt{1+4 a}+4 i
\sqrt{-1-4 a} u_c +3 \sqrt{1+4 a} u_c \right)) +\\\nonumber +6 i a k (-1+u (-1+a
u)) \bar \kappa \epsilon_{\alpha \beta} (\bar B^{\beta}-\bar H^{\beta}  u_c )
(-1+u_c  (-1+a u_c )) \text{ArcTan}\left[\frac{-1+2 a u}{\sqrt{-1-4 a}}\right]
+\\ +6 a k (-1+u (-1+a u))  \bar \kappa \epsilon_{\alpha \beta} (\bar
B^{\beta}-\bar H^{\beta}  u_c ) (-1+u_c  (-1+a u_c ))
\text{ArcTanh}\left[\frac{1-2 a u_c }{\sqrt{1+4 a}}\right])\end{align}

\subsection*{Case $\kappa=0$}

\begin{align}\nonumber B^{\alpha}(u)= \bar B^{\alpha}+\bar H^{\alpha} (u-u_c
)+\\\nonumber+ \frac{1}{6 (2-a) a^3}(-2+a) k \bar u_c \epsilon_{\alpha
\beta}(\frac{2 i (-2+a (-2+3 u))\text{ArcTanh}\left[\frac{1-2 a u}{\sqrt{1+4
a}}\right]}{ (1+4 a)^{3/2} }\times\\\nonumber \times (4 \bar H^{\beta}+a (3 (1+a
(7+2 a (7+a))) \bar B^{\beta}+4 (8+a (2+a) (9+2 a)) \bar H^{\beta}-\\ \nonumber
- 3 (1+a (7+2 a (7+a)))\bar H^{\beta}u_c ))  +\\ \nonumber+\frac{a}{b}( -2 a
\sqrt{1+4 a} (u-u_c ) (6 a (\bar B^{\beta}-\bar H^{\beta} (-8+u))+6 a (\bar
B^{\beta}-\bar H^{\beta}(-8+u)) u_c -8 a \bar H^{\beta} u_c ^2  -\\\nonumber -3
a^2 \left(3 \bar B^{\beta} (-4+u) (1+u_c )+\bar H^{\beta} \left(u (10+3 u)+u
(10+3 u) u_c -2 (-8+u) u_c ^2-4 (7+6 u_c )\right)\right) +\\\nonumber +8\bar
H^{\beta}(1+u_c ) +\\\nonumber +a^4 (\bar H^{\beta} (12+u_c  (18+(-59+12 u (2+3
u)) u_c ))+3 \bar B^{\beta} (2+u_c  (5+12 (-1+u-u_c ) u_c ))) +\\ \nonumber +a^3
9 \bar B^{\beta} \left(4+u (-4+(-4+u_c ) u_c )-u_c  \left(-5+u_c +u_c
^2\right)\right)+\\\nonumber  +a^3 \bar H^{\beta} (29+(23-72 u_c ) u_c +9 u^2
(-4+(-4+u_c ) u_c )+6 u (-4+u_c  (-4+5 u_c )))) +\\ \nonumber + (-2+a (-2+3 u))
(-1+u_c  (-1+a u_c )) ( 2 (4 \bar H^{\beta}+a (3 (1+a (7+2 a (7+a))) \bar
B^{\beta} +\\\nonumber + 4 (8+a (2+a) (9+2 a)) \bar H^{\beta}-3 (1+a (7+2 a
(7+a))) \bar H^{\beta} u_c )) \text{ArcTanh}\left[\frac{1-2 a u_c }{\sqrt{1+4
a}}\right]+\\\nonumber + (1+a) (1+4 a)^{3/2} (-4 \bar H^{\beta}+a (-3 \bar
B^{\beta}-4 \bar H^{\beta}+3 \bar H^{\beta} u_c )) \times \\ \times
(\text{Log}[-1+u (-1+a u)]-\text{Log}[-1+u_c  (-1+a u_c )])        )  ) 
)\end{align}

where $\frac{a}{b} \equiv \frac{1}{(-1-4 a)^{3/2} \left(1+u_c -a u_c ^2\right)}$

\begin{align} \nonumber \nonumber H^{\alpha}_t(u)= \bar H^{\alpha}
\frac{(-1+u) (-1+u (-1+a u))}{(-1+u_c) (-1+u_c (-1+a u_c))}+\frac{(1-u)
\epsilon_{\alpha \beta}}{2 (-1-4 a)^{3/2} a^2}k 
\left(1+u-a u^2\right) \bar u_c \times \\ \nonumber \times (-\frac{1}{(-1+u
(-1+a u)) (-1+u_c  (-1+a u_c ))}2 a \sqrt{1+4 a} (u-u_c )\times \\ \nonumber
\times (4\bar H^{\beta} (1+u) (1+u_c )+a \bar B^{\beta} (3+5 u+5 (1+u) u_c ) 
+\\\nonumber +a\bar H^{\beta}\left(18+u (25+u)+22 u_c +u (25+u) u_c -4 (1+u) u_c
^2\right)   -\\ \nonumber -3a^2 \bar B^{\beta} \left(-5+2 u^2 (1+u_c )+u_c 
(-7+2 u_c )+u (-7+2 (-3+u_c ) u_c )\right)  + \\ \nonumber \bar H^{\beta}a^2
\left(18+24 u_c -22 u_c ^2-6 u^3 (1+u_c )+u (39+5 (7-5 u_c ) u_c )-u^2
\left(1+u_c +u_c ^2\right)\right) + \\ \nonumber +3a^3 \bar B^{\beta} \left(1-8
u_c ^2-u u_c  (3+8 u_c )+2 u^2 (-4+(-4+u_c ) u_c )\right) + \\ \nonumber +\bar
H^{\beta}a^3 (4-4 u_c  (2+5 u_c )+u^2 (-20+(-20+u_c ) u_c )+6 u^3 (-4+(-4+u_c )
u_c ))-\\ \nonumber -\bar H^{\beta}a^3  u (5+7 u_c  (1+5 u_c )) +\bar H^{\beta}
a^4 \left(4 u+4 u_c +6 u u_c +(-2+u (-5+4 u (5+6 u))) u_c ^2\right) +\\
\nonumber+ \bar B^{\beta} a^4(2 u_c +u (2+u_c  (5+24 u u_c )))) +\\\nonumber +2
(4 \bar H^{\beta}+a (3 (1+a (7+2 a (7+a)))\bar B^{\beta} +4 (8+a (2+a) (9+2 a))
\bar H^{\beta} -\\ \nonumber -3 (1+a (7+2 a (7+a))) \bar H^{\beta} u_c )) \left(
\text{ArcTanh}\left[\frac{-1+2 a u}{\sqrt{1+4 a}}\right] +
\text{ArcTanh}\left[\frac{1-2 a u_c }{\sqrt{1+4 a}}\right] \right) + \\
\nonumber +(1+a) (1+4 a)^{3/2} (-4 \bar H^{\beta}+a (-3 \bar B^{\beta}+\bar
H^{\beta} (-4+3 u_c ))) \times \\ \times (\text{Log}[-1+u (-1+a
u)]-\text{Log}[-1+u_c  (-1+a u_c )]))\end{align}

\section*{Acknowledgments}
L.M. would like to thank Francisco Pena-Benitez for his help. This work has been supported by Plan Nacional
de Altas Energías FPA2009-07980, Consolider-Ingenio 2010 CPAN CSD2007-00042,
HEP-HACOS S2009/ESP-247. L.M. has been supported by fellowhship BES-2010-041571. 
  

\end{document}